\documentclass{aa} 
\usepackage{epsfig,times} 
\usepackage{amssymb} 
\usepackage{txfonts} 
\usepackage{graphicx} 
\usepackage{natbib} 
 
\def\be{\begin{equation}} 
\def\ee{\end{equation}} 
\def\beq{\begin{eqnarray}} 
\def\eeq{\end{eqnarray}}

\begin{document} 
 
\title{Magnetic field dissipation in neutron star crusts: 
\\ from magnetars to isolated neutron stars.} 
 
 
\author{J.A.~Pons\inst{1} \and U.~Geppert\inst{1}}
\institute{Departament de F\'{\i}sica Aplicada, Universitat d'Alacant,  
Ap. Correus 99, 03080 Alacant, Spain  
} 
\date{Received...../ Accepted.....} 
 
\abstract 
{We study the non--linear evolution of magnetic fields in  
neutron star crusts with special attention to the influence of the Hall drift. } 
{Our goal is to understand the conditions for fast dissipation due to the 
Hall term in the induction equation. We study the interplay of Ohmic dissipation 
and Hall drift in order to find a timescale for the overall crustal field decay.} 
{We solve numerically the Hall induction equation by means 
of a hybrid method (spectral in angles but finite differences in the radial coordinate). 
The microphysical input consists of the most modern available crustal equation of state, 
composition and electrical conductivities.} 
{We present the first long term simulations of the non--linear magnetic field evolution in realistic 
neutron star crusts with a stratified electron number density and temperature dependent conductivity. 
We show that Hall drift influenced Ohmic dissipation takes place 
in neutron star crusts on a timescale of $10^6$ years. When the initial magnetic field has 
magnetar strength,  the fast Hall drift results in an initial rapid dissipation stage that 
lasts $\sim 10^4$ years. The interplay of the Hall drift with the temporal variation and spatial 
gradient of conductivity tends to favor the displacement of toroidal fields toward the inner 
crust, where stable configurations can last for $\sim 10^6$ years. We show that the thermally 
emitting isolated neutron stars, as the Magnificent Seven, are very likely descendants of neutron 
stars born as magnetars.} 
{} 
 
\keywords{Stars: neutron - Stars: magnetic fields - Stars: evolution} 
\titlerunning{Magnetic field dissipation in neutron star crusts} 
\authorrunning{J.A. Pons \& U. Geppert} 
 
\maketitle 
 
\section{Introduction} 
Since the early days of neutron star (NS) research, when a first estimate of the 
characteristic decay time of its magnetic field was performed \cite{BPP69}, 
our models of magnetic field evolution in 
isolated NSs has become more and more complex.
However, a complete theoretical 
model that explains all observational facts satisfactorily 
does not yet exist.  There is quite general agreement that the magnetic field in 
ultra-magnetized NSs ({\it magnetars}) decays on 
timescales of $10^3$--$10^5$ years \cite{WT06,HL06}. 
Very recently, it has been argued \cite{prl07} that field decay may 
be an effective heating mechanism also in isolated NSs which  
are somewhat less magnetized than {\it standard} magnetars (AXPs and SGRs). 
The observed correlation between surface temperatures and magnetic field strength 
is an evidence for efficient dissipation, and hence Joule heating,
near the surface layers of NSs. 
In addition, the evidence for braking indexes significantly larger than 3,  
which are inferred for an ensemble of a dozen of pulsars in an active 
age of a few $10^5$ years, indicate that epochs of 
rapid decay in relatively young NSs occur (Johnston \& 
Galloway 1999, Geppert \& Rheinhardt 2002). 
 
Another evidence for rapid magnetic field evolution 
in the subsurface crustal layers would be the observation of 
small structures (significantly smaller than the dipolar mode)
close to the surface.  These small scale structures could not 
be survivors from the magnetic field structure acquired by the NS 
at birth; their small scales, in
combination with the low electric conductivity in the outer crust, 
would have erased such field structures on timescales shorter than 1 Myr. 
The existence of small scale structures is, however, 
a necessary condition for the pulsar mechanism to work 
(Ruderman \& Sutherland 1975). In addition, the growing understanding of the physics of 
the drifting sub-pulses supports the idea that small scale crustal field 
structures must be present (Gil et al. 2003, 2006). The 
existence of strong crustal fields in NSs as old as about 1 Myr is 
also necessary to create the non--uniform surface temperature distribution 
observed e.g. in the "Magnificent Seven" (see Pons et al. 2002, 
Schwope et al. 2005; Haberl et al. 2006 for the observational evidence and  
Geppert et al. 2004, 2006; P\'erez--Azor\'{\i}n et al. 2006a, 2006b for the  
theoretical interpretation). All this phenomena can only be understood if  
there is a crustal magnetic field with a strong toroidal component, and
with more structure than the simple dipolar model.
 
On the other hand, population synthesis studies suggest that old pulsars 
show no significant magnetic field decay over their life time, i.e. 
the decay time must be larger than 10 Myr (Hartman et al. 1997;
Regimbau \& de Freitas Pacheco 2001), although the opposite 
conclusion has also been claimed (Gonthier et al. 2004).  
These, at first glance, contradictory facts can be satisfactorily resolved  
by the (quite natural) assumption that the NS magnetic field is  
maintained by two current systems. Long living currents in the
superconducting core support the large scale dipolar field and are responsible 
for the spin down of old pulsars. 
Currents in the crust support the short living part of the field. It decays 
on a timescale of $10^5$--$10^7$ years, depending on the conductivity, 
thickness of the crust, and strength and structure of the initial field.  
Estimates of how fast a core anchored field could be expelled and subsequently 
dissipated in the crust result in characteristic timescales exceeding $100$ Myr 
(Konenkov \& Geppert 2001). 
The dipolar component of the crust is superimposed, outside the NS, 
to the core component. Depending on their
relative strengths, a rapid decay of the crustal dipolar field may or may not 
have observable influence on the pulsar spin down behaviour. 
 
Besides the Ohmic diffusion, which will proceed fast in the outermost 
low--density crustal regions and during the early hot phase of a NS's life, the 
only process that can change the crustal field structure, both 
quantitatively and qualitatively, is the {\it{Hall drift}}. 
Many authors studied during the last two decades the effects of Hall 
drift onto the evolution of magnetic fields in isolated NSs 
(Haensel et al. 1990; Goldreich \& Reisenegger 1992; Muslimov 1994; Naito \& 
Kojima 1994; Urpin \& Shalybkov 1995; Shalybkov \& Urpin 1997;
Vainshtein et al. 2000; Rheinhardt \& Geppert 2002; Geppert \& Rheinhardt 2002;
Hollerbach \& Ruediger 2002;  Geppert et al. 2003; Rheinhardt et al. 
2004; Cumming et al. 2004). By use of the analogy of the Hall induction 
equation with the vorticity equation of an incompressible liquid,
Goldreich \& Reisenegger (1992) developed the idea of the 
{\it{Hall cascade}}. It transfers magnetic energy from the 
largest to smaller scales until a critical scale length below which Ohmic  
decay becomes dominant is reached. Vainshtein et al. (2000) considered for the 
first time the Hall drift in a stratified NS crust. In the presence of a density profile, 
Hall currents are able to create current sheets, which are places where
very efficient dissipation occurs. This effect is dramatic if current sheets 
are located just below the surface, where the conductivity is lower,
but even if the drift is directed towards  
the highly conductive inner crust, the small scale of the locally intense field 
causes a significantly faster dissipation of magnetic energy than the 
purely Ohmic diffusion estimate. 
In summary, this MHD--like process (considering the crust as a one--component plasma) 
is by itself energy conserving. However, it affects the magnetic field by two 
inherent tendencies: the creation of small scale field structures by transferring 
magnetic energy from the initial large scale field and, 
the drift of growing structures towards a region where current sheets are created.

Another interesting effect of the Hall drift was proposed by
Rheinhardt \& Geppert (2002). They show, by means of a stability analysis of the 
linearized Hall induction equation, that the 
transfer of energy from the large scale (background) field to the smaller scale modes may proceed  
in a non--local way in the momentum space, resulting in a Hall instability.  
Cumming et al. (2004) showed that the growth rate of this instability depends on the 
shear in the velocity of the electrons whose current supports the background field. As they 
pointed out it is unclear whether the {\it {Hall instability}} is relevant 
for the field evolution in the crust, since the Hall cascade may proceed sufficiently fast to 
fill in the intermediate scales between the large scale initial field and the unstable growing small scale ones. 
 
In the present study we intend to consider the field evolution as realistic as possible. 
Thus, we will solve numerically the non--linear Hall induction equation 
in the crustal region, using a crustal density and conductivity profile 
according to the state-of-the-art microphysical input. We start with initial 
field configurations which consist both of poloidal and toroidal field 
components that fulfill boundary conditions at the superconducting 
core and at the surface. We will follow the cooling history of a NS, starting with an initially 
hot NS where the magnetization parameter is relatively small and the field evolution in 
the crust is almost completely determined by Ohmic diffusion. As the NS cools down, the 
magnetization parameter increases, thereby gradually 
enhancing the relative importance of the Hall drift. 
Our aim is to reach a better understanding of the effects that Hall drift may have 
on the crustal field evolution and its consequences for the rotational and thermal evolution 
of isolated NSs.
 
The paper is organized as follows. In Sect. 2 we present the Hall induction equation where the 
magnetic field is represented in terms of its poloidal and toroidal components which are expanded 
in a series of spherical harmonics. 
Sect. 3 is devoted to purely toroidal fields. For this case we show that 
the induction equation can be written in a form similar to the Burgers equation. 
This allows us to discuss the importance of the field--dependent drift in a clear way. 
In Sect. 4 we detail the input microphysics,  
the NS model, and the initial conditions. In Sect. 5 we present the results for different initial 
field configurations.  Finally, in Sect. 6, we discuss our main conclusions and give an outlook 
to future improvements. 
 
\section{Basic Equations and Formalism} 
 
In the crystallized crust of NSs, where convective motions of the 
conductive material play no role, the evolution of the magnetic field is 
governed by the Hall--induction equation 
\beq 
\frac{\partial\vec B}{\partial t}= - \frac{c^2}{4\pi}{\nabla \times}\left(\frac{1}{\sigma} 
\left\{\nabla \times \vec{B} + \omega_B \tau [(\nabla \times \vec{B}) \times \vec b ]\right\} \right), 
\label{Hallind} 
\eeq 
where $\vec b$ is the unit vector in the direction of the magnetic field $\vec b= \vec{B}/B$, 
with $B$ being the magnetic field strength, $\tau$ is the relaxation time of the electrons 
and $\omega_B=eB/m^*_e c$ is the electron 
cyclotron frequency. Here, $\sigma=e^2 n_e \tau/m^*_e$ is the electrical conductivity parallel 
to the magnetic field , and $n_e$ the electron number density.
The Hall drift term (the term proportional to $\omega_B\tau$) at the r.h.s. 
of Eq. (\ref{Hallind}) is a 
consequence of the Lorentz force acting on the electrons.
The tensor components of the electric conductivity are derived in the relaxation 
time approximation (Yakovlev \& Shalybkov 1991).  
If the {\it magnetization parameter} $\omega_B\tau$ 
exceeds unity significantly, the Hall drift dominates, which 
results in a very different field evolution from the purely Ohmic case.
A large magnetization 
parameter, typically $\approx 1000$ during some stages \cite{GR02,paper2}, 
strongly suppresses the electric conductivity perpendicular to the 
magnetic field.  
 
The up to now limited success to solve Eq. (\ref{Hallind}) has been 
restricted to magnetization parameter not exceeding 200  
\cite{HR02} 
\footnote {Their parameter $R_B$ is equivalent to our $\omega_B\tau$ 
calculated for a fixed, initial magnetic field strength $B_0$.}. 
These numerical limitations guided us to use an alternative approach, 
since fully spectral codes have systematically unsurmountable problems to deal 
with structures where discontinuities or very large gradients of the  
variables appear. This will happen, for example, if current sheets develop as  
a consequence of the Hall drift. The next section will describe in detail 
the reason why spectral methods may not be well suited to deal with this problem. 
We have decided to employ a semi--spectral method which describes the angular part of 
the field by spherical harmonics but uses a spatial difference scheme in 
radial direction. 
 
The formalism that uses the representation of 
fields by their poloidal and toroidal parts and their expansion in a series of 
spherical harmonics was developed by R{\"a}dler (1973). 
R{\"a}dler's formalism allows to transform the (vector) induction equation 
into two scalar equations. In this formalism, the magnetic field is decomposed in 
poloidal and toroidal components: 
\beq 
\vec{B}=\vec{B}_{\rm pol} + \vec{B}_{\rm tor}~. 
\eeq 
Hereafter, we follow the notation of Geppert \& Wiebicke 
(1991). The two components are described by two functions, 
$\Phi (r,\theta,\varphi,t)$ and $\Psi (r,\theta,\varphi,t)$,  
where $r,\theta$, and $\varphi$ are the usual spherical coordinates.  
Explicitly, the components of the field are given by 
\beq 
\vec{B}_{\rm pol} &=& \nabla \times \vec{A}~, \quad \quad 
\vec{A} = - \vec{r} \times \nabla \Phi ~,
\nonumber \\ 
\vec{B}_{\rm pol} &=&= -\vec{r} ~\nabla^2 \Phi + \nabla 
\left( \frac{\partial (r\Phi)}{\partial r}\right) ~,
\nonumber \\ 
\vec{B}_{\rm tor} &=& - \vec{r} \times \nabla \Psi ~.
\label{bfield} 
\eeq 
This formulation has the advantage that it automatically fulfills the divergence 
condition ($\nabla \cdot \vec{B} = 0$). Inserting the expressions of Eq. (\ref{bfield}) into 
Eq. (\ref{Hallind}) we arrive at two partial differential equations  describing the 
evolution of the poloidal and of the toroidal part of the magnetic field. 
\beq 
\frac{\partial{\Phi}}{\partial t} &=& \frac{1}{\hat{\sigma}}~\nabla^2 \Phi + D ~,
\nonumber \\ 
\frac{\partial {\Psi}}{\partial t} &=& \frac{1}{\hat{\sigma}}~ 
\left( \nabla^2 \Psi -\frac{1}{r} \frac{\partial \log \hat{\sigma}}{\partial r} 
\frac{\partial (r\Psi)}{\partial r} \right) + C~. 
\label{PhiPsi} 
\eeq 
Here, $\hat{\sigma}= 4\pi\sigma/c^2$ and $D$ and $C$ stand for the nonlinear 
terms describing the Hall drift and coupling both components 
(poloidal and toroidal) of the magnetic field. We have assumed that the
conductivity depends only on the radial coordinate, and is independent of
the magnetic field strength.
 
After expanding the functions $\Phi$ and $\Psi$ in a series 
of spherical harmonics 
\beq 
\Phi = \frac{1}{r} \sum_{n,m} \Phi_{nm}(r,t) Y_{nm}(\theta,\phi)~, 
\nonumber \\ 
\Psi = \frac{1}{r} \sum_{n,m} \Psi_{nm}(r,t) Y_{nm}(\theta,\phi)~,
\label{expans} 
\eeq 
where $n=1,\ldots,n_{\rm max}$ and  $m=-n,\ldots,+n$.
The vector potential, as well as the poloidal and toroidal parts of the 
magnetic field, can be written as 
\beq 
\vec{A} &=& - \left( \frac{1}{r} \sum_{n,m} \Phi_{nm} 
\frac{d Y_{nm}}{d \theta} \right)~\vec{e}_{\phi}~, 
\nonumber \\ 
\vec{B}_{\rm pol} &=& \frac{1}{r^2} \sum_{n,m} n(n+1) \Phi_{nm} Y_{nm}~ \vec{e}_{r} 
+ \frac{1}{r} \sum_{n,m} \frac{\partial \Phi_{nm}}{\partial r} 
\frac{d Y_{nm}}{d \theta}~\vec{e}_{\theta}~, 
\nonumber \\ 
\vec{B}_{\rm tor} &=& - \left( \frac{1}{r} \sum_{n,m} \Psi_{nm} 
\frac{d Y_{nm}}{d \theta} \right)~\vec{e}_{\phi}~, 
\label{bpoltor} 
\eeq 
while the corresponding components of the current density, given by 
$\vec{J}= \frac{c}{4\pi} \nabla \times \vec{B}$, is 
\beq 
\frac{4\pi}{c} \vec{J} &=& \frac{1}{r^2} \sum_{n,m} n(n+1) \Psi_{nm} Y_{nm} ~\vec{e}_{r} 
+ \frac{1}{r} \sum_{n,m} \frac{\partial \Psi_{nm}}{\partial r} 
\frac{d Y_{nm}}{d \theta}~\vec{e}_{\theta} 
\nonumber \\ 
&+& \frac{1}{r} \sum_{n,m} \left[ 
\frac{\partial^2 \Phi_{nm}}{\partial r^2} - \frac{n(n+1)}{r^2}\Phi_{nm} 
\right] \frac{d Y_{nm}}{d \theta} ~\vec{e}_{\phi}. 
\label{currents} 
\eeq 
Finally, by inserting the expansions of Eq. (\ref{bpoltor}) into Eq. (\ref{Hallind}), 
we arrive to an infinite set of partial differential equations: 
 
\beq 
\frac{\partial {\Phi_{nm}}}{\partial t} &=& \frac{1}{\hat{\sigma}}~ 
\left( \frac{\partial^2 \Phi_{nm}}{\partial r^2} - \frac{n(n+1)}{r^2} 
\Phi_{nm} \right) + D_{nm} ~,
\nonumber \\ 
\frac{\partial {\Psi_{nm}}}{\partial t} &=& \frac{1}{\hat{\sigma}}~ 
\left(  \frac{\partial^2 \Psi_{nm}}{\partial r^2} - \frac{n(n+1)}{r^2} \Psi_{nm} 
- \frac{1}{\hat{\sigma}}\frac{\partial \hat{\sigma}}{\partial r} 
\frac{\partial \Psi_{nm}}{\partial r} \right) + C_{nm} ~. 
\label{PhiPsi_ind} 
\eeq 
In this paper we will restrict ourselves to axially symmetric field 
configurations, i.e., the index $m=0$ and we will drop it henceforth. Following 
the derivation of Geppert \& Wiebicke (1991) the nonlinear coupling terms are 
\beq 
D_{n}= \frac{\hat{\tau}}{\hat{\sigma} r^2} \sum_{k,k'} I^{(2)}\left( 
\frac{\partial}{\partial r}\Phi_k\Psi_{k'} - 
\frac{\partial}{\partial r}\Psi_k\Phi_{k'}\right)\;, 
\label{Ind_Phi} 
\eeq 
\beq 
C_{n} &=& \sum_{k,k'} I^{(2)}\frac{\partial}{\partial r} 
\left(\frac{\hat{\tau}}{\hat{\sigma} r^2}\left[ 
\Psi_k\Psi_{k'} + \Phi_k^{(1)}\Phi_{k'}\right]\right) + 
\nonumber \\ 
& & \frac{\hat{\tau}}{\hat{\sigma} r^2} \sum_{k,k'} I^{(3)} 
\left[\Psi_k\frac{\partial}{\partial r}\Psi_{k'} + 
\frac{\partial}{\partial r}\Phi_k\Phi_{k'}^{(1)}\right]\;, 
\label{Ind_Psi} 
\eeq 
where 
\beq 
\Phi_k^{(1)} = \left(\frac{\partial^2}{\partial r^2} 
-\frac{k(k+1)}{r^2}\right)\Phi_k~.  
\eeq 
Above, $\hat{\tau}$ = $\omega_B\tau/B$, and $I^{(2)},I^{(3)}$ are expressions
that contain the Clebsch--Gordan coefficients which reflect the coupling properties of the field 
modes with different multipolarity (see Eq. (60) in Geppert \& Wiebicke (1991)). 
 
Using the orthonormality properties of the spherical harmonics, 
the volume integrated magnetic energy can be calculated as 
\beq 
\frac{1}{8 \pi} \int &dV& B^2 = 
\nonumber \\ 
\frac{1}{8 \pi} \int &dr& \sum_{n} n(n+1) \left[ n(n+1) \left( \frac{\Phi_n}{r}\right)^2 
+ \left(\frac{d\Phi_n}{dr}\right)^2 
+ {\Psi_n}^2 \right]\;. 
\eeq 
As a criterion for the magnetic energy conservation we will check 
that the equality 
\beq 
\frac{1}{8 \pi} \frac{d}{dt} \int B^2~dV = 
- \int \frac{J^2}{\sigma} \; dV~, 
\eeq 
is satisfied during the evolution within a certain tolerance (typically less 
than $10^{-3}$). The integral on the l.h.s is evaluated over the volume occupied
by the magnetic field (including the vacuum exterior region), while the integral
on the r.h.s. is calculated over the region where currents exist (the crust).
 
\subsection{Outer Boundary Conditions} 
\subsubsection{Outside Vacuum}
 
Since we consider in this study realistic crusts with finite  
electric conductivity,  surface currents are excluded.  
This means that we require all components of the magnetic field to be  
continuous across the NS surface, i.e. 
that the scalar fields $\Phi_n$ and $\Psi_n$, and the 
derivative $\frac{\partial \Phi_n}{\partial r}$ 
are continuous through the outer boundary (see also R{\"a}dler 1973). 
 
The external vacuum solution of a NS magnetic field  is 
determined by $\nabla \times \vec{B}=0$, $\nabla \cdot \vec{B}= 0$ and
the boundary conditions. From Eq. (\ref{currents}),
the vanishing curl condition leads to 
\beq 
\frac{1}{r} \sum_{n} \left[  
\frac{\partial^2 \Phi_{n}}{\partial r^2} - \frac{n(n+1)}{r^2}\Phi_{n} 
\right] \frac{d Y_{n}}{d \theta} \vec{e}_{\phi} = 0\;\;{\rm at}\;\; r \ge R, 
\eeq 
which can be expressed as $\Delta \Phi =0$, where $\Delta$ is the Laplacian. 
The index $m=0$ in $Y_{nm}$ has been omitted for simplicity. The only physical  
solution of this equation is $\Phi_n=a_nr^{-n}$. Therefore, the requirement  
of continuity across the surface results in  
\beq 
\frac{\partial \Phi_{n}}{\partial r} = -\frac{n}{R}\Phi_n\;\; {\rm at}\;\; r=R. 
\label{OBC_Phi} 
\eeq 
 
Because the poloidal current must also vanish  
in vacuum, we can derive another general boundary condition 
\beq 
\sum_n\left[\frac{\Psi_n}{r\sin{\theta}} 
\frac{\partial (\sin{\theta}\frac{\partial Y_n}{\partial \theta})} 
{\partial \theta}\vec{e_r} 
+ \frac{1}{r} 
\frac{\partial{(r\Psi_n\frac{\partial Y_n}{\partial \theta}})}{\partial r} 
\vec{e_{\theta}}\right] = 0\;\;{\rm for}\;\; r\ge R~. 
\eeq 
The existence of surface currents may affect the condition on the 
$\theta$--component of the poloidal current density, but there cannot 
exist radial currents penetrating into the vacuum. 
Thus the outer boundary condition for the toroidal field is simply 
\beq 
\Psi_{n} =0 \;\; {\rm at}\;\; r=R. 
\label{OBC_Psi} 
\eeq 
 
In general, the poloidal tangential $\theta$--component of the  
current density has to vanish at the surface only if the electric conductivity  
vanishes there. Then, according to Ohm's law, any finite tangential current  
density would cause an infinite tangential electrical field which is in  
contradiction to the energy conservation guaranteed by Maxwell's equations.  
Therefore, in the case of $\sigma=0$ at $r=R$, any solution of the induction  
equation which fulfills the boundary condition $\Psi_{n} =0 \;\; {\rm at}\;\, 
r=R$ will be characterized by vanishing tangential surface currents,  
i.e. $ \Psi_{n}=\frac{\partial \Psi_n}{\partial r}=0 \;\; {\rm at}\;\; r=R$.

\subsection{Inner Boundary Conditions} 
 
The inner boundary conditions are determined by the transition from normal to 
superconducting matter at the crust--core interface $r=R_i$. The Meissner--Ochsenfeld 
effect demands that the normal component of the magnetic field has to vanish at 
$r=R_i$. Continuity of the tangential component of 
the electric field together 
with Ohm's law enforces that component to vanish at $r=R_i$ because otherwise  
the infinite electric conductivity would cause infinite tangential current  
densities, finally destroying the superconducting state. 
 
For a spherically symmetric NS, the normal component  
of the magnetic field is its $r-$component as given by Eq. (\ref{bpoltor}).  
Thus, the inner boundary condition for the poloidal field is 
\beq  
\Phi_n = 0\;\;\; {\rm{at}}\;\;\; r = R_i \;\;. 
\label{IBC_Phi_1} 
\eeq 
 
The tangential component of the electric field  
consists of  $\theta-$ and $\varphi-$components. For the $\varphi-$component we find, 
after some algebra: 
\beq 
E_{\varphi}&=&\sum_{n,n'}\left[ \frac{n(n+1)}{r^3} \Psi_n  
\frac{\partial \Phi_{n'}}{\partial r} Y_n 
\frac{\partial Y_{n'}}{\partial \theta}  
- \frac{n'(n'+1)}{r^3} \frac{\partial \Psi_{n}}{\partial r}\Phi_{n'} 
\frac{\partial Y_n}{\partial \theta} Y_{n'}\right] 
\nonumber \\ 
&+& \frac{1}{r\omega_{\rm B}\tau}\sum_n\left(\frac{\partial^2 \Phi_{n}}{\partial r^2} 
-\frac{n(n+1)}{r^2}\Phi_n\right)\frac{\partial Y_n}{\partial \theta}\;. 
\eeq 
 
This electric field component has to vanish at the surface of  
the superconducting core. Since $\Phi_n=0$ at $r=R_i$ , this condition  
reads 
\beq 
\sum_{n,n'}\left[n(n+1) \Psi_n  
\frac{\partial \Phi_{n'}}{\partial r} Y_n 
\frac{\partial Y_{n'}}{\partial \theta} \right] 
+ \frac{r^2}{\omega_{\rm B}\tau}\sum_n\left(\frac{\partial^2 \Phi_{n}}{\partial r^2} 
\frac{\partial Y_n}{\partial \theta}\right)\; =\; 0\;. 
\eeq 
 
This condition is obviously not suited to find an inner boundary  
condition for the toroidal field. Therefore, we have to consider the  
$\theta-$component of the electric field which consists of three contributions: 
\beq  
E_{\theta} = \left[ (\nabla \times \vec{B}_{\rm tor}) \times  
\vec{B}_{\rm tor} +  (\nabla \times \vec{B}_{\rm pol} ) \times  \vec{B}_{\rm pol} +   
\frac{1}{\omega_B \tau}(\nabla \times \vec{B}_{\rm tor}) \right]_{\theta}
\eeq 
 
Using again the expressions of Eq. (\ref{bpoltor})  
and taking again into account that $\Phi_n(r=R_i)=0$, the term  
$\left(\nabla \times \vec{B}_{\rm pol}\right) \times \vec{B}_{\rm pol}$ 
vanishes and it remains from the condition $E_{\theta}=0$ 
at the inner boundary that
\beq 
\frac{1}{\omega_{\rm B}\tau}\sum_n \frac{\partial \Psi_n}{\partial r} 
\frac{\partial Y_n}{\partial \theta} + 
\frac{1}{r^2}\sum_{n,n'} n(n+1) \Psi_n\Psi_{n'}Y_{n'} 
\frac{\partial Y_n}{\partial \theta} = 0\; . 
\eeq 
Multiplying both sides with $\frac{\partial Y_l^{\ast}} 
{\partial \theta}$ and integrating over the solid angle the orthonormality of 
the spherical harmonics returns  
\beq 
\frac{1}{\omega_{\rm B}\tau}\frac{\partial \Psi_n}{\partial r}=-\frac{1}{r^2} 
\sum_{k,k'} I^{(2)} \Psi_k\Psi_{k'}\;\;.
\label{IBC_Psi} 
\eeq 
Hollerbach \& R{\"u}diger (2002) applied the above boundary condition in 
the limit of $\omega_B \tau \rightarrow \infty$, avoiding thereby the difficulties the 
non--linearity will cause. We use the general form of Eq.~(\ref{IBC_Psi}), since
during the cooling process of the NS, there certainly will be
periods during which it is not justified to neglect the dissipative term. 
 
\begin{figure} 
\resizebox{\hsize}{!}{\includegraphics{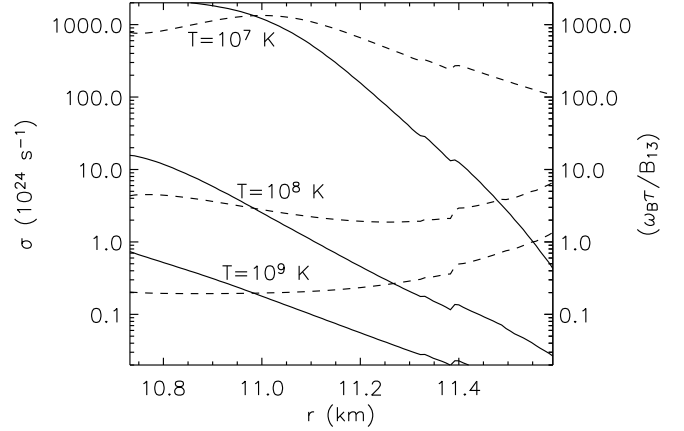}} 
\caption{Radial profiles of electrical conductivity $\sigma$ (solid lines), 
and the magnetization parameter normalized to a fiducial magnetic field 
of $B_{13}=10^{13}$ G (dashed lines), for three different temperatures. 
The crust is assumed to be isothermal. 
} 
\label{fig1} 
\end{figure} 
 
\section{Comments on the evolution of purely toroidal fields} 
 
From Eqs. (\ref{PhiPsi}), (\ref{Ind_Phi}), and (\ref{Ind_Psi}) it is readily 
seen that an initial purely toroidal field ($\Phi=0$ at $t=0$) will not 
develop a poloidal part and does remain purely toroidal. 
For clarity, we will use in this section the $\varphi$--component 
of the magnetic field ($B_\varphi$) as a variable. 
In axial symmetry, it is related to $\Psi$ through  
\beq 
\vec{B}_{\rm tor} \equiv B_\varphi \vec{e_{\varphi}}  
=  - \frac{\partial \Psi}{\partial \theta} \vec{e_{\varphi}}~. 
\eeq 
In order to make evident the effect of the Hall drift we will use cylindrical coordinates 
($R, \varphi, z$).  In the Appendix A we have written the general form 
of the induction equation in cylindrical coordinates, with a  
decomposition of the magnetic field as used by Hollerbach \& R\"udiger (2002). 
When considering the case of purely toroidal fields, neglecting for simplicity the dissipative term 
(limit of strong magnetization), and assuming constant electron density $n_e$, the 
induction equation is reduced to a single 
evolution equation for the variable $B_\varphi$: 
\beq 
\frac{\partial B_\varphi}{\partial t} = -
\frac{\hat{\tau}}{\hat{\sigma}} \vec{e_{\varphi}} \cdot 
\nabla \times [(\nabla \times \vec{B}_{\rm tor}) \times \vec{B}_{\rm tor}] 
= \frac{c}{4 \pi e n_e} 
\frac{2B_\varphi}{R} \frac{\partial B_\varphi}{\partial z}\;\;, 
\label{Burgers} 
\eeq 
where we have used that ${\hat{\tau}}/{\hat{\sigma}}={c}/{4 \pi e n_e}$.  

In this form, the Hall--dominated induction equation has the form of the 
inviscid Burgers equation with a wave velocity that depends on the variable $B_\varphi$, 
and on the coordinate $R$. This has been pointed out by Reisennegger et al. (2005)
who also concluded that it leads to the formation of current sheets.
In this way, the formulation illustrates very clearly that the originally  
(in the limit of weak field) purely parabolic diffusion equation changes its character 
to hyperbolic when the Hall term dominates. The analogy between the induction 
equation and the Burgers equation was discussed before by Vainshtein et al. (2000) but there  
are a few more qualitative differences we should notice.  
Vainshtein et al. (2000) derived the analogy with the Burgers equation  
in Cartesian coordinates, assuming no dependence of the field on the $z$-coordinate.  
In this particular case, it is necessary to consider a stratified medium in order to  
have the Hall drift leading to the formation  of current sheets. Our result is more 
general: even in the uniform density case, the geometry of a NS crust (axially 
symmetric toroidal field  in a conducting {\it spherical shell}) leads to a Burgers-like 
equation that does not admit stationary solutions. Depending 
on the sign of the gradient, the field will drift {\it vertically} in one or 
the other direction, leading to the formation of current sheets either at 
the surface, or at the crust-core interface.  
The purely vertical drift acting within a spherical shell (the crust)  
will result inevitably in fast dissipation locally, wherever current sheets are formed.
Moreover, when we consider NS models with a stratified
electron number density, the gradient in $n_e$ through the NS crust may additionally  
support the creation of current sheets (mainly close to the crust--core interface).
 
But there is yet another important problem related to the change from parabolic to hyperbolic 
character of the induction equation for purely toroidal fields. It is well known that the 
solutions of the inviscid Burgers equation develop 
discontinuities or, if a small viscous term is present, steep gradients. This is 
probably the reason why spectral methods have always failed when trying to 
solve numerically the induction equation for large magnetization parameters. 
The system evolves naturally to form a discontinuity or sharp gradient 
in $B_{\varphi}$ on a characteristic timescale which can be shown to be
$\approx \tau_{\rm Hall}$ (the Hall timescale is defined below in Eq.~(\ref{tau_Hall}))
\beq
\tau_{\rm shock} = \left(  \frac{c}{2\pi e n_e R} \frac{\partial B_\varphi}{\partial z} \right)^{-1} =
\frac{R \hat{\sigma}}{2 \hat{\tau}} \left( \frac{\partial B_\varphi}{\partial z} \right)^{-1}~,
\eeq
which is of the order of $\approx 10^6-10^7$ yr for $B=10^{13}$ G and typical lengthscale of 1 km. 
However, it can be much shorter for stronger fields and small scale structures generated by the Hall cascade. 
A similar timescale characterizes the typical travel time of a magnetic perturbation to reach either 
the surface or the core--crust boundary, 
\beq
\tau_{\rm travel} = \left(  \frac{c}{2\pi e n_e R} \frac{B_\varphi}{d} \right)^{-1},
\eeq
where $d$ denotes the distance between the initial position of the perturbation in the crust and the 
surface or the crust--core interface.
The formation of a shock by the compression of the field against the inner 
or outer boundaries will cause unavoidable numerical instabilities  
and/or the Gibb's phenomenon if one tries to solve the problem by means of spectral methods. 
Note that, locally and temporarily, the three characteristic timescales 
($\tau_{\rm Hall},\tau_{\rm shock},$ and $\tau_{\rm travel}$)
may coincide or differ by orders of magnitude, because they are proportional to the second, 
first, and zeroth derivative of the field strength, respectively. 
These characteristic timescales become different once small-scale structure
or current sheets appear, creating small length scales in addition to the
larger crust size.
This makes any reasoning based on those timescales questionable; 
only a serious numerical study of the Hall drift can yield an idea about its 
effects on the crustal magnetic field evolution.

For the simple case of a Burgers-like equation with constant $n_e$, (see Eq. (\ref{Burgers})), 
we have checked  that upwind methods, specifically designed to deal with hyperbolic equations, 
work very well in regimes $\omega_B \tau \rightarrow \infty$, in which a spectral method fails. 
Thus, the numerical problems observed with spectral codes for $\omega_B \tau \ga 100$ are most 
likely caused by an intrinsic limitation of the numerical approach. 
In the general case, having both poloidal and toroidal field components in a stratified spherical shell, 
the field evolution is not so simple. Then, the equations are strongly coupled, they have both a 
parabolic and a hyperbolic part and it is difficult to guess what is the best strategy to solve them. 
It is not the scope of this paper to give the final answer about the best numerical technique. 
We intend to point out in this section that there are deep unavoidable reasons that lead to 
unsurmountable problems in many cases. 
For the rest of the paper, we will focus on simulations with realistic NS  
models that can be handled by our hybrid method. We take advantage of a  
seldom occasion: the more realistic model causes less numerical problems than the  
constant density model. The reason is that current sheets are faster smoothed out in  
the shell layers just below the surface, because their electric conductivity is orders  of magnitudes 
smaller than in the inner crust. In addition, in the long run, the toroidal field seems to find a 
quasi-equilibrium configuration in which the effect of the gradient of conductivity is 
counterbalancing the Hall drift. Such an equilibrium can not be established for non-stratified crustal models.
 
\begin{table} 
\caption{Description of initial models, differing in the 
initial field strength and relative sign of the poloidal and 
toroidal components. All initial poloidal fields are purely
dipolar ($n=1$). The magnetic field is in units of $10^{13}$ G.} 
\begin{tabular}{lccc} 
\hline 
\hline\noalign{\smallskip} 
{Model} & $B_{\rm pol}$   & $B_{\rm tor}$ & {Multipole}\\ 
{} & ($10^{13}$ G) & ($10^{13}$ G) & {(toroidal)}\\ 
\hline 
A & 10 & -100 & $n=2$ \\ 
B & 10 &  100 & $n=2$ \\ 
C & 10 & -100 & $n=1$ \\ 
D &  1 &  -10 & $n=2$ \\ 
E & 20 & -200 & $n=2$ \\ 
F & 50 &  -50 & $n=1$ \\ 
G &100 & -100 & $n=2$ \\ 
\hline\noalign{\smallskip} 
\end{tabular} 
\label{models} 
\end{table} 
 
\begin{figure*} 
\resizebox{\hsize}{!}{\includegraphics{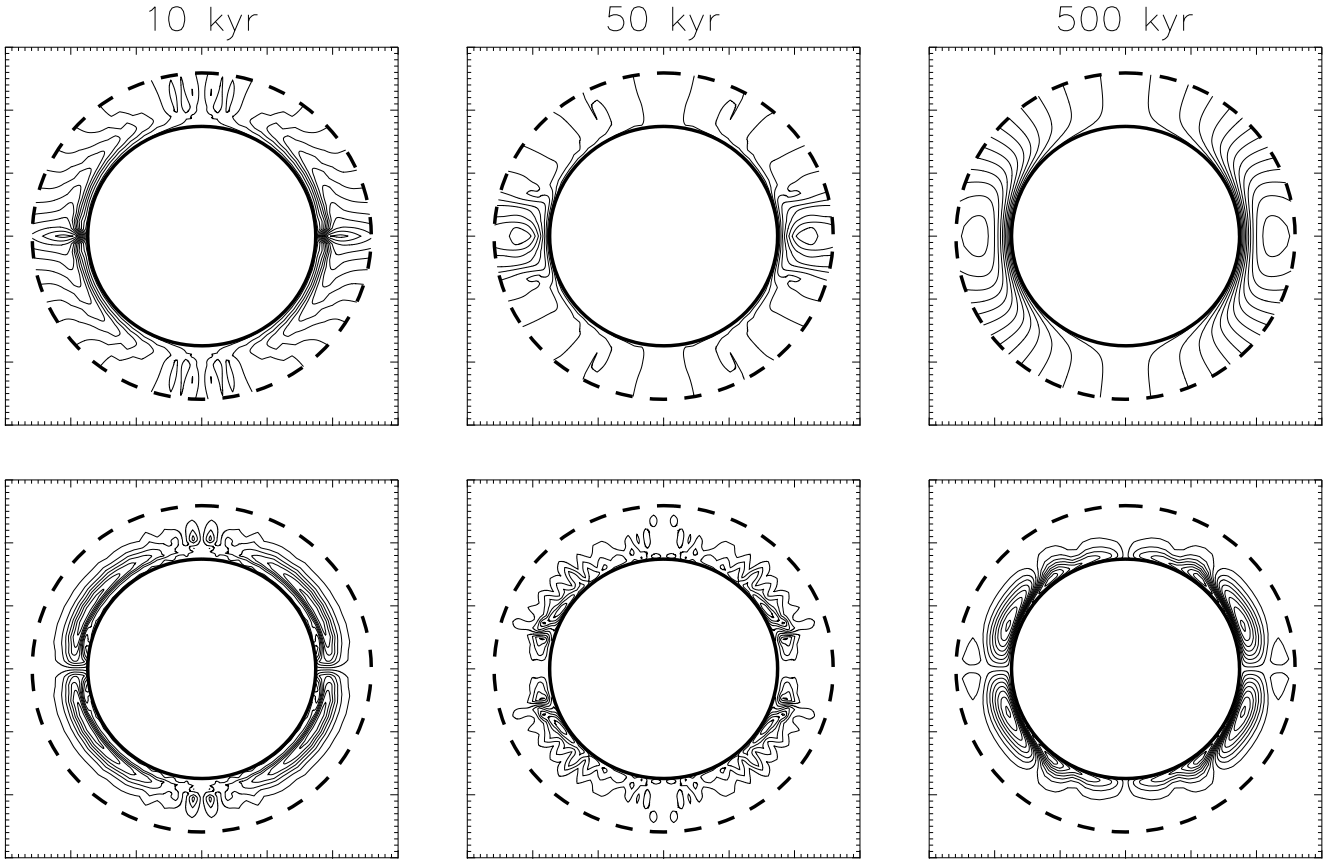}} 
\caption{Top: Poloidal magnetic field lines for three  
different evolutionary stages of the magnetic 
field configuration initially described by model A. 
($t=10, 50,$ and $500$ kyr). Bottom: contours of constant  
toroidal field strength ($B_\varphi$). 
The crustal shell has been stretched a factor of 4 for clarity. 
} 
\label{fig2} 
\end{figure*} 
\section{The NS model and the initial conditions.} 
 
Our aim is to study the global evolution of the magnetic 
field in isolated NSs. The Hall drift occurs both in the fluid core 
and in the solidified crust. While the effect of the Hall drift in the core is 
less obvious and may proceed on a timescale of the order of the Hubble time 
(insert $L_5=10\; {\rm and}\; \rho \approx 10\rho_{\rm nuc}$ into Eq. (61) of 
Goldreich \& Reisenegger (1992)), its effect can be crucial in the crustal field. 
  
In order to build the background NS model we have used a modern  
Skyrme-type equation of state (EOS) at zero temperature 
describing both, the NS crust and the liquid core, based on  
the effective nuclear interaction SLy \cite{DH91}. 
The low density EOS (below the neutron drip point) employed is 
that of Baym et al. (1971). With this EOS, we have built a NS 
model having a radius of about 11.7 km and a mass of 1.28 $M_\odot$. The central 
density is 8.83$\times 10^{14}$ g/cm$^3$ and its crust (from $\rho \approx
10^{10}$ g/cm$^3$ to $\approx 10^{14}$ g/cm$^3$) extends from 10.7 to 11.6 km. 
In Fig. \ref{fig1} we show radial profiles of the electric conductivity  
$\sigma$ (solid lines) and $\omega_B \tau/B_{13}$ (dashed lines), where 
$B_{13}$ is $B$ in units of $10^{13}$ G, for three different temperatures. 
We have assumed a constant impurity concentration parameter
$Q = {n_{\rm imp}(Z_{\rm imp}-Z)^{2}}/{n_{i}}$ of $Q=10^{-2}$. Here, 
$n_{\rm imp}$ and $n_{i}$ are the impurity and ion particle densities, 
respectively, and $Z_{\rm imp}$ is the charge number of the impurities.
We have compared simulations with $Q=10^{-2}$ and $Q=10^{-4}$ without finding
significant differences. A high impurity content could lead to
even faster dissipation \cite{jones}.
The figure illustrates the fact that the electric conductivity varies by  
$3 - 4$ orders of magnitude within the crust and depends strongly  
on the temperature. The magnetization parameter scales linearly with $B$. 
For the fiducial field of $10^{13}$ G, it is of the order of unity for 
a temperature of $10^9$ K but can become as large as $1000$ 
as the star cools down. For magnetar field strength which both at the 
surface (radial poloidal component) and within the crust (meridional
poloidal and toroidal components) may well exceed $B_{13}=10$, the
magnetization parameter can reach locally values in excess of $10000$. 

From Fig. \ref{fig1} one can also read the relevant timescales of the problem.
The Ohmic timescale is
\be
\tau_{\rm Ohm}=\frac{4\pi \sigma \lambda^2}{c^2} = \hat{\sigma} \lambda^2
\ee 
where $\lambda$ is the typical magnetic field length-scale. Inserting some typical numbers, we get
\be
\tau_{\rm Ohm}= 4.4 \left(\frac{\sigma }{10^{24} {\rm s}^{-1}}\right)
\left(\frac{\lambda}{\rm km}\right)^2~ 10^6~{\rm years}
\ee 
On the other hand, the Hall timescale is 
\be
\tau_{\rm Hall}=\frac{4\pi e n_e \lambda^2}{c B}~.
\label{tau_Hall}
\ee
The ratio of the Ohmic to Hall
timescale is simply given by $\omega_B \tau$.
From Fig. \ref{fig1} we can infer that, if the temperature of a NS's crust varies 
between $10^9$ and $10^8$ K during the first million years of its life, the average
Ohmic timescale in the crust is $\approx 1$ Myr. For magnetized NSs, the ratio 
of the Ohmic to Hall timescale is approximately
\be
\frac{\tau_{\rm Ohm}}{\tau_{\rm Hall}} =(1-10)\times B_{13}~. 
\ee

\begin{figure*} 
\resizebox{\hsize}{!}{\includegraphics{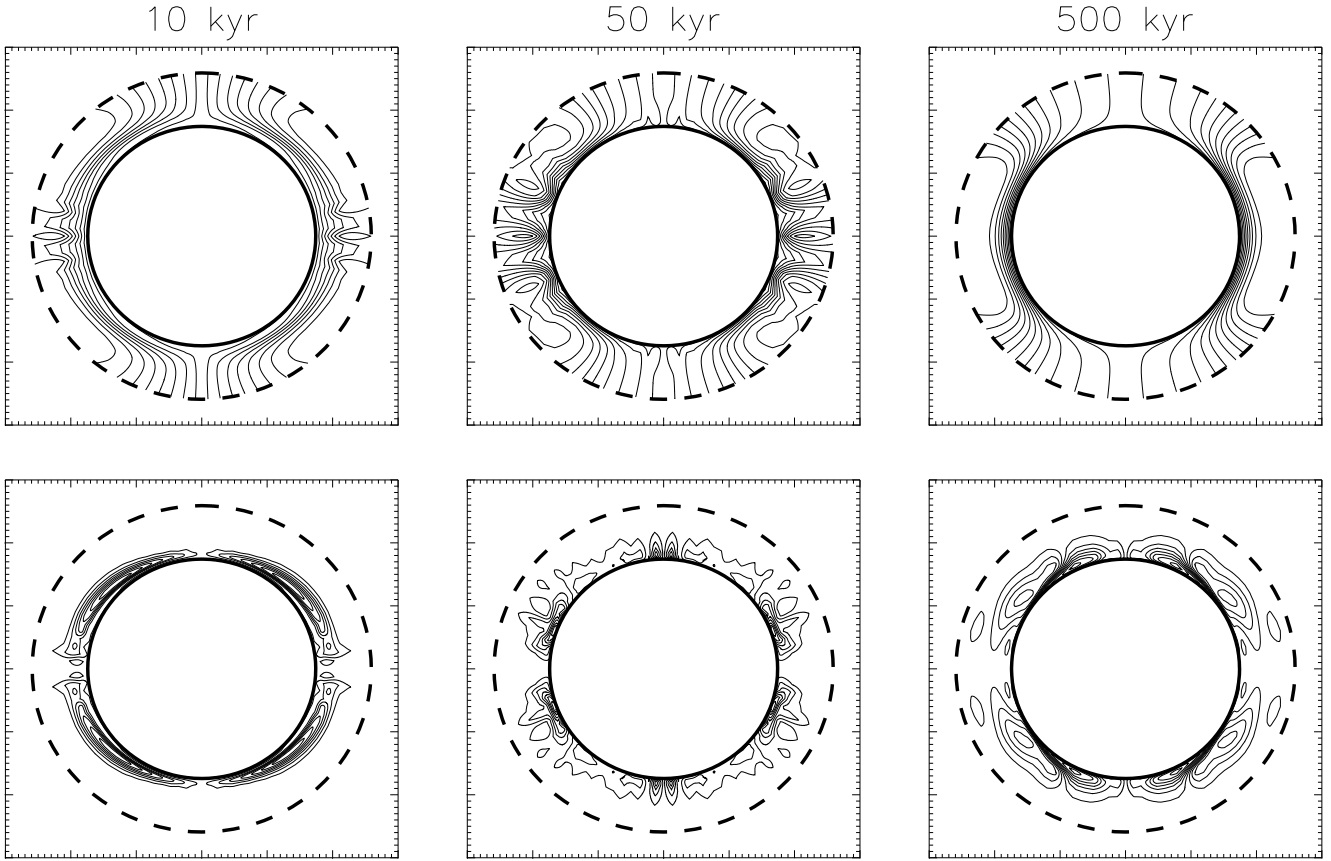}} 
\caption{Same as Figure \ref{fig2} but for model B. 
} 
\label{fig3} 
\end{figure*} 
 
Note, however, that such averaged timescales are of very restricted use
to characterize the field evolution in NS crusts, since both the density and
the magnetic field vary over many orders of magnitude there.
There is no doubt, then, that in any circumstances there is magnetic
field decay in the crust of young NSs. A different issue is whether or not
this effect is observable when studying populations of older NSs. Because of
the strong temperature dependence of the electrical conductivity, when a NS's crust
cools below $10^7$ K, typically one million years after birth, the Ohmic
dissipation time increases significantly, and no rapid field decay can be expected
after that age.

For this reason, in this paper we want to focus on the initial evolution 
of magnetic fields in relatively young NS. In order to mimic the cooling process of a NS, 
we have started with a crust at $T=10^9$ K, a typical value after formation 
of the crust, at most within hours after birth, and we force the temperature of
the isothermal crust to vary according to 
\beq 
T(t) = 10^9 (1 + 10^{6} t_6)^{-1/6} 
\eeq 
where $t_6$ is the NS age in $10^6$ years. This approximation is valid during the neutrino 
cooling era if only modified URCA processes are operating (Page et al. 2006). 
It is not the purpose of this paper to discuss how the thermal and magnetic 
field evolution are coupled in detail, but this simple approximation is  
sufficient to capture the main effect: as the NS's crust cools (from $10^9$ to 
$10^8$ K in about 1 Myr), via the decrease of the electron relaxation time ($\tau$), 
the Hall term becomes more and more important. When a significant part of the crustal field 
is dissipated and/or it has approached a force free configuration, its decay 
continues on a much longer Ohmic timescale. 
We have not included effects of temperature anisotropies within the 
crust, although they may be important. This will be addressed in detail in 
future works.  
 
The other microphysical input needed for performing the simulations is 
the electrical conductivity and the magnetization parameter. Since, 
as the temperature drops and the magnetic field evolves, the magnetization parameter 
and the electrical conductivities vary, 
at each time step we recompute the value of the relaxation time 
and the electrical conductivities by using the public code 
developed by A. Potekhin (1999) \footnote{{\tt www.ioffe.rssi.ru/astro/conduct/condmag.html}}. 
Despite it causes some numerical efforts, we have chosen to be as realistic 
as possible and use this state-of-the-art microphysical ingredients to account 
for the effect of composition stratifications.
 
\begin{figure*} 
\resizebox{\hsize}{!}{\includegraphics{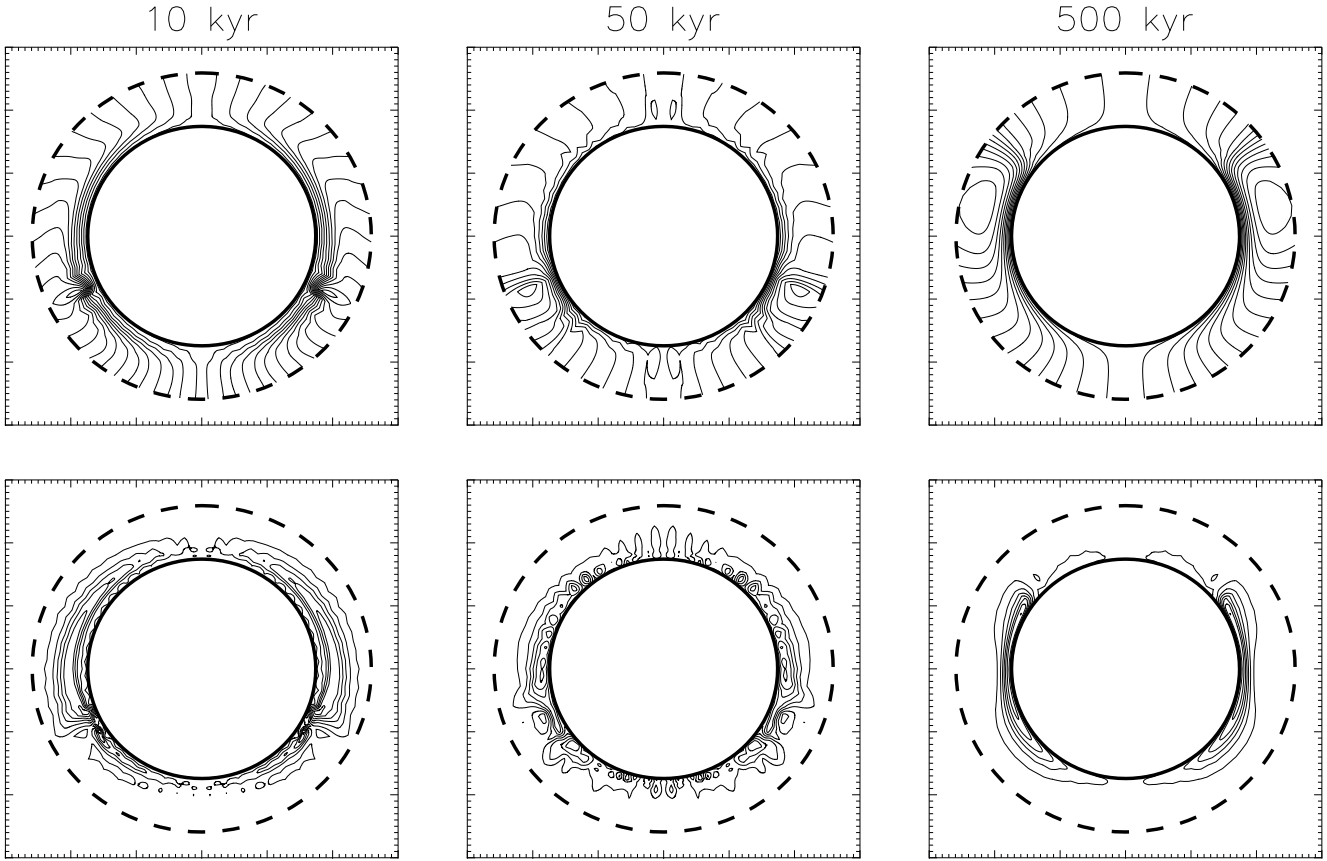}} 
\caption{Same as Figure \ref{fig2} but for model C. 
} 
\label{fig4} 
\end{figure*} 

\subsection{Initial magnetic field configuration.} 
 
Little is known with certainty about the initial magnetic field  
strength and structure. Recently, Braithwaite (2006) and Geppert \& Rheinhardt  
(2006) have shown that sufficiently rapid rotation can stabilize dipolar  
toroidal and poloidal fields of magnetar strength against  
MHD instabilities occurring immediately after completion of the proto-NS phase. 
Only preliminary results are available about the stability of  
magnetic field structures consisting of a poloidal--toroidal mixture.  
Our conclusion about the origin of strong toroidal crustal fields that probably exist in 
magnetars and quite certainly exist in thermally emitting NSs with a 
highly non--uniform surface temperature distribution (Geppert et al. 2004, 2006;  
P\'erez--Azor\'{\i}n et al. 2006a, 2006b) is, therefore, that 
they have probably been generated at birth. Later, they were stabilized 
against Tayler instabilities, and have been frozen into the solid crust when it forms. 
Alternatively, the presence of sufficiently strong temperature gradients both close to the 
crust--core interface and in the degenerate surface layers 
(Wiebicke \& Geppert 1995) are able to convert thermal  
into magnetic energy very effectively via a thermoelectric instability. 
In this case, the magnetic energy will be predominantly stored in toroidal field structures. 
 
We have considered quite different initial structures: purely 
poloidal and purely toroidal ones, with initial dipolar and quadrupolar 
modes, and several mixed initial fields where the ratio between the energies 
stored in the toroidal and poloidal field parts is varied. 
For the poloidal field, our initial configurations are chosen to 
be free Ohmic decaying modes (strictly speaking, pure decay modes 
for constant conductivity profiles) satisfying the boundary conditions. 
For a given angular multipole ($n$), the radial eigenfunctions  
can be written as linear combination 
of spherical Bessel functions of the first (${\rm j}_n$) and second (${\rm n}_n$) kind 
\beq 
\Phi_n(r) = a~ r~ {\rm j}_n(\mu r) + b~ r~ {\rm n}_n(\mu r) 
\eeq 
where $\mu$ is the radial wavenumber. 
The boundary conditions fix the wavenumber $\mu$ and the ratio $b/a$. 
For the models presented in this paper,  
the overall normalization factor has been chosen to fix the radial 
component at the pole of the dipolar ($n=1$) component to the values
of $B_{pol}$ listed in Table \ref{models}.
 
The initial toroidal field is given by 
\beq 
\Psi_l(r) = c [(r-R_{i})(r-R)]^2 
\eeq 
where the constant $c$ is determined by the maximum value of $B_\varphi$. 
In Table 1 we summarize the models employed in this work, differing in the initial 
values and relative sign of the poloidal and toroidal components. 
 
\section{Results and discussion.} 
 
We present now the results of our numerical simulations of the field  
evolution in the crust and its dependence on the initial magnetic field structure and strength.
We will restrict this presentation to the selection of initial models described 
in Table \ref{models}, although we have performed a number of different 
simulations with a variety of initial configurations. In particular, we 
have reproduced some of the toy models (constant $\hat{\sigma}$ and $\hat{\tau}$)
found in the literature \citep{HR02} to test our code. We saw a good qualitative 
agreement with only minor quantitative differences. We have also performed simulations 
of the evolution of purely toroidal fields, which remain toroidal forever, for 
testing purposes. For conciseness, we will discuss now only realistic models 
of magnetized NSs. 
 
\begin{figure} 
\resizebox{\hsize}{!}{\includegraphics{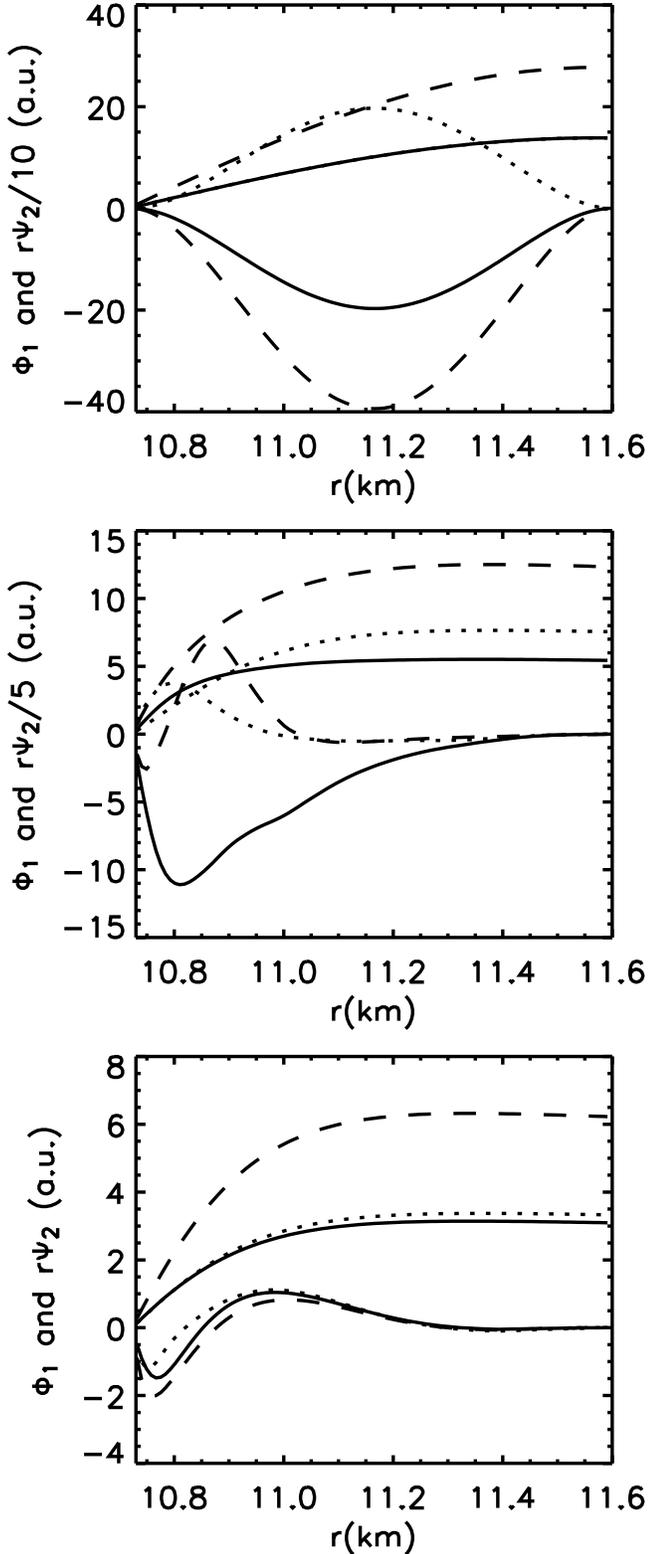}} 
\caption{Radial profiles of $\Phi_1$ and $r \Psi_2$ (in arbitrary units). 
These quantities are proportional to the poloidal (dipolar) and toroidal 
(quadrupolar) components, respectively (see Eq. \ref{bpoltor}). 
The upper panel shows the initial model, while the central and bottom 
panel correspond to $t=20$ kyr and $t=500$ kyr, respectively. 
The different line styles stand for models A (solid line), 
B (dotted line), and E (dashed line). Notice the different normalization
of $r \Psi_2$ in the upper and middle panel (a factor of 10 and 5, respectively).
} 
\label{fig5} 
\end{figure} 

In Fig. \ref{fig2}, \ref{fig3}, and \ref{fig4} we present the evolution of models $A$, $B$ and $C$. 
For three representative ages, we show both the poloidal magnetic field lines (top panels) and 
the contours of constant toroidal field strength (bottom panels). The crustal shell has 
been stretched by a factor of 4 for clarity. 
Models $A$ and $B$ differ only in the relative sign of its toroidal component 
which is for both models initially quadrupolar ($n=2$), while model $C$ has an initially dipolar ($n=1$) 
toroidal field with the same radial profile as model $A$. 
In all cases the models have the same maximum toroidal field strength.  
The different sign of the toroidal field affects the direction of the Hall drift (toward north or south). 
The other important effect is due to the gradient of the electric conductivity,
which always causes the tendency to displace field lines towards the inner crust. 

For Model A, we see in the central panel how the Hall drift compresses magnetic field lines 
near the equator,  while the results for Model B show the opposite tendency. 
Model C, due to its dominant initial dipolar structure, shows a global displacement of the 
magnetic field  towards the south pole, that results in a fast dissipation in the 
high resistivity surface region near the south pole. During that phase, the interplay 
of the poloidal and toroidal field modes is very intense. The magnetic field is dragged
and twisted, thereby creating current sheets. At these sites of very 
efficient dissipation the toroidal field weakens rapidly and, after becoming about equally 
strong as the poloidal field, the latter tends to return into its original position. 
This sequence of twisting and stretching and subsequent release of the twisted 
field characterizes this epoch by its oscillatory behaviour.
However, looking at the right panels, we see that after about half a million years, 
all poloidal fields have a very similar appearance. 
The strongly dissipated, and now weaker, toroidal components,
become more regular than in the Hall phase, and they are dominated by the $n=2$ or $n=3$ modes. 
\footnote{The black/white figures
presented here are snapshots and yield only a vague impression 
of the evolution of the magnetic field with Hall drift and Ohmic 
diffusion, everything coupled with the cooling of the crust.
We advice the reader to look at the movies, available under 
{\tt http://www.dfa.ua.es/UNS06/movies.html}, 
where the evolution is visualized in much more detail.}
 
The main conclusion from these results is that the initial magnetic 
field configuration largely determines the early 
evolution ($< 10^5$ yr). On the long run, however, there seems 
to exist a clear tendency to establish a more stable configuration, consisting 
of a dipolar poloidal component in combination with higher order  
toroidal field modes. This fact can be better understood when looking at the 
evolution of the radial profiles of the dominant modes, as shown in Fig. \ref{fig5}. 
At early times, say $t=20$ kyr, the interplay between the 
toroidal and poloidal components results in a different evolution of $\Phi_1$, 
although the shape is qualitatively similar. 
After half a million years, the poloidal field is clearly dominant, 
and very similar for all models. The toroidal field is weaker 
(in real units, about a factor of 2-3 smaller than the poloidal field) 
and concentrated towards the inner crust. It is interesting to notice 
that also $\Psi_2$ is rapidly driven towards a similar shape in all cases, 
indicating that after the fast initial transient the gradient of 
the conductivity determines a sort of {\it quasi-equilibrium} field. 
 
\begin{figure} 
\resizebox{\hsize}{!}{\includegraphics{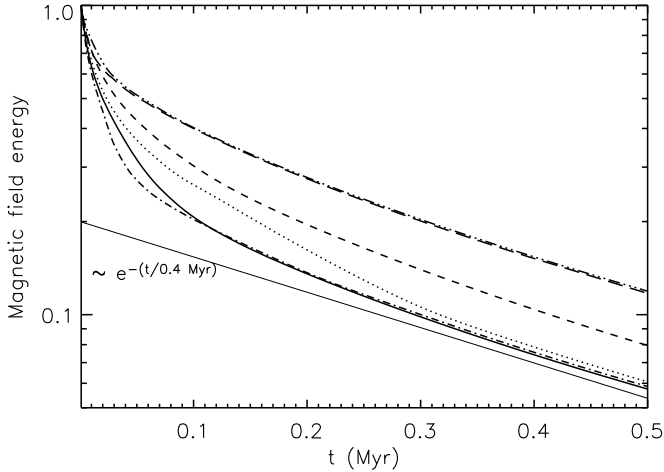}} 
\caption{Evolution of the total energy stored in the magnetic field 
,normalized to the initial value,
for models A (solid), B (dotted), C (short dashes), E (dot-dash), 
F (triple dot-dash), and G(long dashes). The thin line shows, for comparison, an
exponential decay with a timescale of $\approx 0.4$ Myr.
} 
\label{fig6} 
\end{figure} 
 
It is also interesting to look at the evolution of the total magnetic energy 
(Fig. \ref{fig6}) to understand the different evolutionary stages that the 
NS passes through. Firstly, it is remarkable how all models have a very similar evolution, 
and how they converge at late times towards a similar exponential decay. 
For all models, the magnetic energy has been dissipated by a factor 10 to 20 after 0.5 Myr.
Secondly, we can clearly differentiate between the initial fast decay, 
in which the field can dissipate one half of its energy in only $\sim 10^4$ yr. This 
is caused by the much smaller initial conductivity at higher temperatures combined with the 
effects of the Hall drift. During the subsequent stage, the Ohmic dissipation timescale 
(about half a million years) is almost identical for all models. 
Obviously, the first stage is  of great importance for models with initially stronger fields, 
i.e. it will play an important role for NSs born as magnetars. 
This seems to be the only period, before the photon cooling era, when the Hall drift will 
qualitatively affect the field evolution. 
In general, the models in which the initial ratio of toroidal to poloidal field is small 
(models F,G) decay slower than those with large ratios (models A,B,C,E). Among the latter,
the stronger the initial field, the more efficient is the dissipation during the Hall phase.

The Ohmic dissipation rate appears to be faster than what has been usually discussed in 
the literature. The reason is that we are simulating for the first time the evolution of NS 
models with a realistic profile of electron density, composition, conductivity, etc., 
{\it including} the coupling between different field modes through the non--linear Hall term.  
The effect of the Hall drift is twofold: it causes the transfer of magnetic energy to smaller scale 
modes that decay faster and, in some cases, the displacement of the magnetic field to regions of 
higher resistivity where it is rapidly dissipated. 

\begin{figure} 
\resizebox{\hsize}{!}{\includegraphics{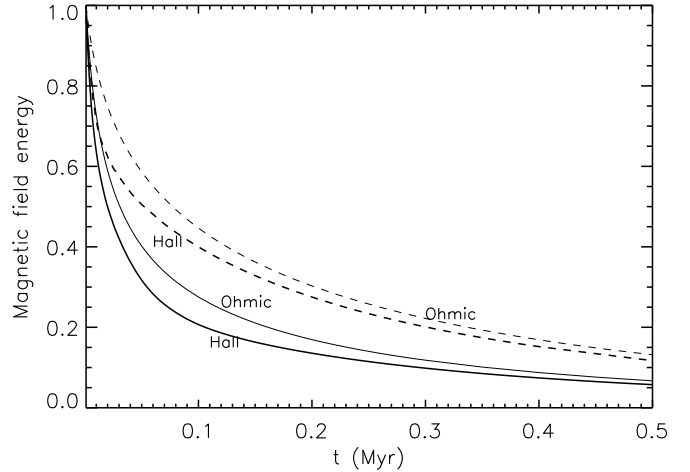}} 
\caption{Evolution of the total energy stored in the magnetic field 
for models A (solid) and G (dashed) compared with the same initial configurations
evolving only by Ohmic decay (thin lines). The difference between thick and thin
lines is due to the nonlinear Hall terms, that result in a faster initial 
dissipation.} 
\label{fig7} 
\end{figure}

In Fig. \ref{fig7} we compare the evolution of the total magnetic energy in
models A and G (thick dashes and solid lines, respectively) with the hypothetical
evolution of the same initial configuration without the Hall drift (by setting $\hat{\tau} = 0$).
After the initial phase (about 50 kyr) during which the models that include the Hall 
drift dissipate energy about $10-20 \%$ faster, the rate of energy dissipation becomes 
approximately the same. This makes evident that
the initial departure of the exponential decay is partially caused by Hall
drift, but also by the relatively rapid cooling of the crust that results in a 
significant time-dependence of the conductivity and of the magnetization parameter. 
Other cooling models, for example fast cooling due to the activation of direct URCA processes, 
may result in an initially different evolution. However, we think that on a secular timescale of about 
0.1 Myr the system readjusts itself and the dissipation of magnetic field is mostly controlled by the 
Ohmic decay. In a future work we plan to couple a multidimensional cooling code with the magnetic 
field evolution to study different cooling scenarios.
 
\begin{figure*} 
\resizebox{\hsize}{!}{\includegraphics{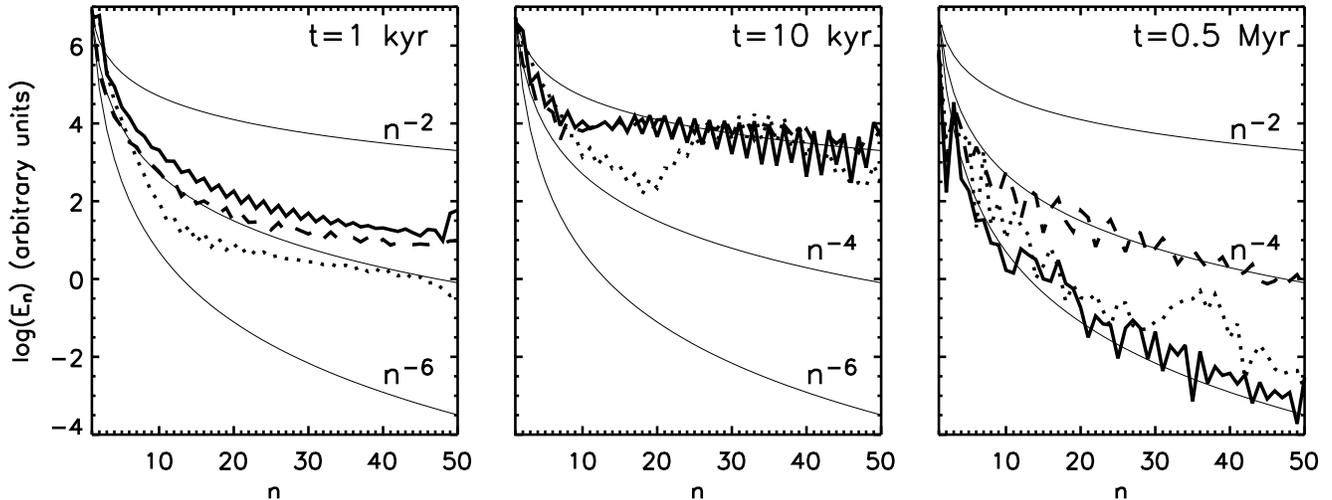}} 
\caption{Power spectrum for models 
A (solid lines), B (dots), and C (dashes) at three different epochs.
For comparison, we also show three different power laws (thin lines).
} 
\label{spect} 
\end{figure*} 

To finish the discussion in this section, we present in Fig. \ref{spect} the power spectra 
for models A, B, and C at three different moments of the evolution. In the left panel, 
which corresponds to $t=1$ kyr, we see how the Hall cascade is filling very quickly 
the shorter wavelength modes (initially only $n=1$ and $n=2$ modes
existed). At $t=10$ kyr, which corresponds roughly to the Hall timescale,
the Hall cascade and perhaps the Hall instability have filled out all
large wavenumber modes and approximately saturates following a $n^{-2}$
power law. This situation is kept for about 50 kyr (not shown in the figure),
until the field has been dissipated by a significant amount, and the Hall term 
begun to lose its importance with
respect to the regular Ohmic dissipation term. After half a million years
the power spectrum is much steeper ($ \propto n^{-6}$), an indication
that the Hall drift has lost its influence on the crustal field evolution.
 
\begin{figure}
\resizebox{\hsize}{!}{\includegraphics{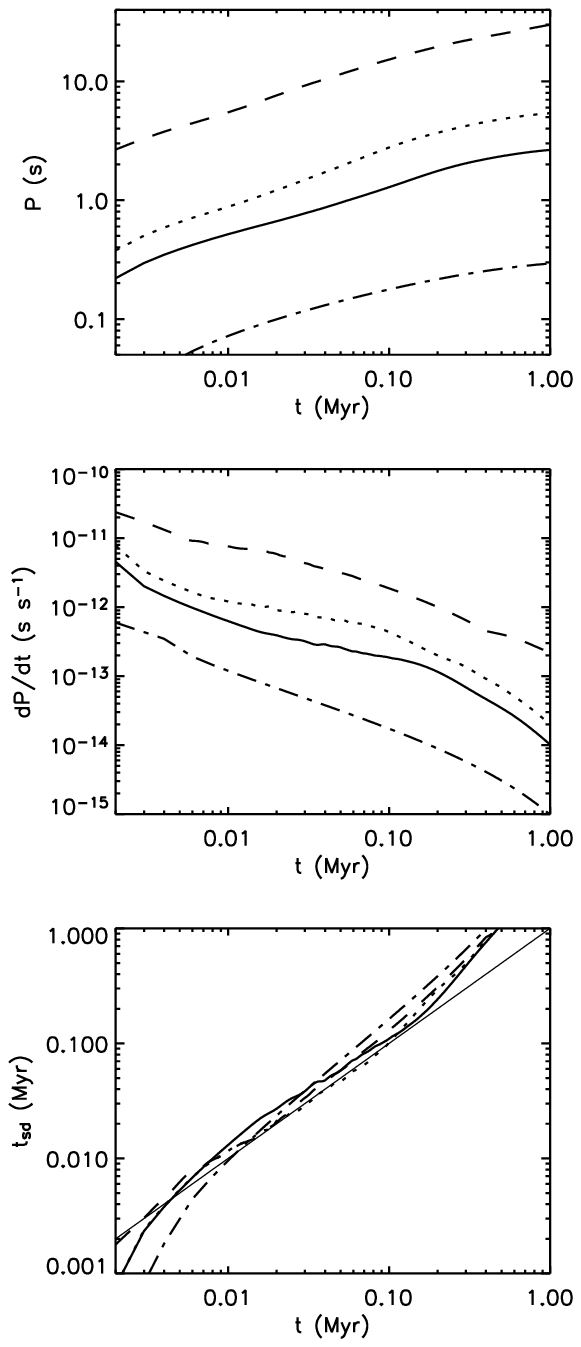}} 
\caption{Evolution of the period (top), period derivative (middle) and
spin-down age (bottom) in four selected models with different initial
dipolar field strengths. The line-styles correspond to models 
A (solid lines), D (dash-dotted line), E (dots), and G (dashes).
A typical NS born as a magnetar would correspond to model G, while a
typical pulsar such as Geminga would show the behaviour of model D.
} 
\label{period} 
\end{figure} 
 
\subsection{Spin down evolution.} 
 
Since the evolving crustal field may also have effects on the rotational evolution 
of the NS, we study the differences of the spin down behaviour between different
models. In Fig. \ref{period} we show the evolution of the period (top), period derivative 
(middle) and spin-down age ($t_{\rm sd}=P/(2\dot{P})$) (bottom) in four selected 
models with different initial 
dipolar field strengths. The line-styles correspond to models 
A (solid lines), D (dash-dotted line), E (dots), and G (dashes).
A typical NS born as a magnetar would correspond to model G or, perhaps, E, while a
typical pulsar such as Geminga would show the behaviour of model D.

\begin{figure} 
\resizebox{\hsize}{!}{\includegraphics{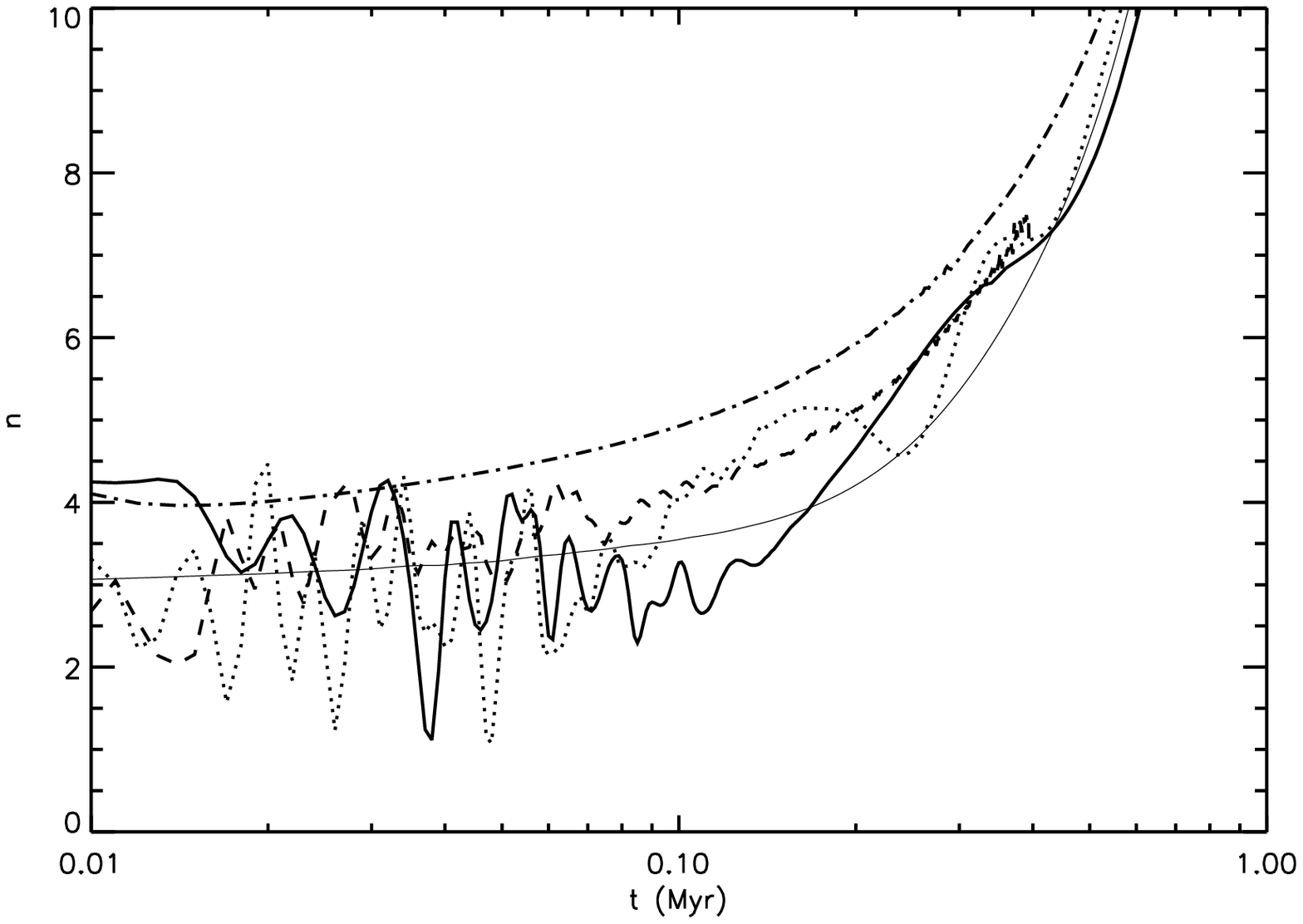}} 
\caption{Evolution of the braking index $n=\nu \ddot{\nu}/\dot{\nu}^2$ for the same models as
in Fig. \ref{period}.
} 
\label{brak} 
\end{figure} 

Notice that a minimum initial poloidal field of about $2\times 10^{14}$ G is
required to explain the large rotation periods ($5-10$ s) of thermally emitting 
isolated NSs as the Magnificent Seven. This may be another indication towards a 
common evolutionary path in which some INSs were born as magnetars and their 
magnetic field has decreased one order of magnitude during their lifetimes.
The evidence for crustal field decay presented here has also
implications for estimates of the ages of old NSs. Pulsar spin down is thought 
to generally follow the prediction of the vacuum dipole model, which gives $\dot\nu\propto
B^2\nu^3$, where $\nu$ is the spin rate. If the birth spin rate far
exceeds the present spin rate and $B$ is constant, the age in this
model is given by $t_{\rm sd}$. This expression is
used as the standard estimate of a pulsar's age. If the field is
decaying according to our simulations, however, the relationship
between the true age and $t_{\rm sd}$ given by the dipole model is
shown in the bottom panel of Fig. \ref{period}. It shows that
$t_{\rm sd}$ seriously {\em overestimates} the age for NS older than $10^5$ years.
This effect helps to reconcile the
observed discrepancy between the spin-down ages and independent
measures of the ages of some isolated NSs.

Another interesting feature is the rapid variation of the dipolar poloidal
component during the Hall epoch. Of course, the total magnetic energy always decreases,
but the $n=1$ poloidal component may exhibit an oscillatory behavior. This is a result 
of the efficient energy transfer between the large scale toroidal and the dipolar 
poloidal field modes, i.e. an genuine effect of the Hall drift. 
Assuming that only the $n=1$ component contributes to spin down the star,
the oscillations of the dipolar poloidal surface field affect 
the braking index $n=\nu \ddot{\nu}/\dot{\nu}^2$, as shown in Fig. \ref{brak}. 
During the first $\sim 10^5$ years of their lives, sufficiently magnetized NSs 
will show a quite erratic variation of the braking index, that may reach any value 
from 1.5 to 4. The oscillations of the braking index continue even until 
$\sim 5\times10^5$ years, when the initial toroidal field is stronger, as seen for 
model E which has an initial maximum toroidal field of $2\times 10^{15}$ G. 
At late times, once Ohmic dissipation is controlling the evolution
and the dipolar magnetic field decreases steadily, the braking index gradually increases and 
might reach very high values. This process will be inverted after the crustal field is dissipated 
almost completely and the rotational evolution of the NS is determined by the much longer timescale 
of the core field expulsion. In that period, the braking index will approach its canonical 
(dipolar) value of $3$ again. Our results for the braking index evolution coincide very well 
with the observed indices for middle aged pulsars (see Geppert \& Rheinhardt g002).
It can be shown that the braking index satisfies the relation
\beq
n=3-2\frac{\dot{B}P}{B\dot{P}} \approx 3 + 4 \frac{t_{\rm sd}}{\tau_B},
\eeq
where we have introduced a typical magnetic field decay time $\tau_B=B/\dot{B}$ to
derive the latter equality. The thin solid line in Fig. \ref{brak} corresponds to this
approximation with $\tau_B=0.8$ Myr. Notice that Fig. \ref{fig7} shows the evolution of
the total magnetic energy ($\propto B^2$), so that the decay time for the magnetic field
is twice the value quoted in Fig. \ref{fig7}.

It should be noted that after $\approx 1$ Myr neutrino emission no longer 
controls the thermal evolution of a NSs, and our results do not apply.
Moreover, the low temperatures reached at that time would increase significantly
the conductivity (see Fig. \ref{fig1}). Then, the purely Ohmic decay will proceed
much slower, but also the magnetization parameter will increase, which may result
in another {\it Hall stage} during the photon cooling era. In future work, 
we plan to extend our study to longer times with consistent temperature evolution.
In addition, one always must keep in mind that the magnetic field component
supported by currents and in the superconducting NS core would be dominant after
the crustal field has been dissipated.  This, however, is a completely different 
scenario that cannot be analyzed with our presently available tools.
 
\subsection{The Hall instability ?} 
We have seen in our numerical studies some hints for the occurrence of the Hall instability, i.e.,
for the rapid, non--local in the momentum space, energy transfer from the initial large scale field 
modes into much smaller ones. Related spectral features can be seen e.g. in Fig.~\ref{spect}, where 
local maxima appear. However, at the present level of energy conservation ($99.9\%$) and our limiting angular 
resolution ($n_{\rm max}=120$, although most of our runs were performed with $n_{\rm max}=50$) we can not 
clearly distinguish the truncation effects from the onset of the Hall instability. Therefore, we decided 
to postpone the study of the Hall instability in the NS crust to future work.

\section{Conclusions} 
The above presented results show that magnetic fields maintained by currents 
circulating in the crust of NSs are strongly rearranged
and do decay significantly during the first million years of a NS's life.
In addition to purely Ohmic decay, which is faster in the first thousands of
years when the electrical conductivity is relatively large due to the high crustal
temperature, we find that the Hall drift may contribute noticeably to
accelerate the dissipation of magnetic fields. 
For typical field strengths of $10^{14}$ G, we observe a {\it Hall drift dominated} 
stage followed by purely Ohmic decay proceeding in a timescale of the order of 1 Myr. 
Depending on the strength and structure of the initial magnetic field, this Hall phase 
lasts a few $10^3$--$10^4$ years and it is characterized by an intense exchange of 
magnetic energy between the poloidal and toroidal components of the field and by the 
redistribution of magnetic field energy between different scales.
It can be expected that such rearrangements and the relatively rapid field 
decay have observational consequences, as can be observed in magnetars.
Of course, whether this first phase plays an important role and how distinctive 
it is depends on the initial field strength and structure and how fast the NS cools.  
If a NS begins its life as a magnetar its external dipolar field is $> 10^{14}$ G. 
Within the crust, however, the internal magnetic field may locally exceed that value by about 
one order of magnitude. Thus, even the expected initial high crustal temperatures 
of $\ga 10^9$K, which cause a relative small electron  relaxation time ($\tau$), 
cannot avoid that $\omega_B\tau \gg 1$ in a large fraction of the crust volume. 

The toroidal part of the field is specially affected by the Hall drift. 
There are two main effects acting upon the toroidal fields: it is globally displaced toward the 
inner crust because of the almost everywhere negative conductivity gradient and, depending on 
the relative sign with respect to the poloidal component, it tends to move vertically toward
one or the other magnetic pole. After the Hall stage, during which the toroidal field is strongly
rearranged and dissipated, the long term evolution seems to select, generally, a predominantly 
quadrupolar/octupolar structure concentrated in the inner crust and with tendency to be stronger 
close to the poles. This multipolar structure will determine the surface temperature distribution 
of middle aged NSs, that could be more complex than previously thought.
Such complex field structures and the local deposition of energy by Joule heating 
favor surface temperature distributions characterized not only
by two hot polar caps, but for example by a hot equatorial belt, as has been probably 
seen in RX J0720.4-3125 (Haberl et al. 2006, P\'erez--Azor\'{\i}n et al. 2006b) 
or can be inferred from the light curve of RBS 1223 (Schwope et al. 2005).

Though the toroidal part of the crustal field undergoes a spectacular dissipation and 
rearrangement, the coupling between both parts also affects the evolution of the poloidal field,
which is responsible for the spin down of NSs. Thus,
the complex interaction of Hall drift and Ohmic dissipation is also reflected in the 
temporal behaviour of the braking index and, in principle, accessible for observations.  
Our models show that the braking index of young ($\la 10^5$ years), magnetized NSs
exhibits a quite wild variation, and can reach any value from 1.5 to 4. 
After the Hall stage, when Ohmic dissipation controls the evolution
and the dipolar magnetic field decreases steadily, the braking index gradually increases and 
might reach very high values. This process will cease after the crustal field is dissipated 
almost completely, or when the conductivity increases as the NS cools down.
During the photon cooling era, the braking index will approach its canonical 
(dipolar) value of $3$ again. 

If the initial magnetic field is too weak ($\la 10^{12}$G) for the Hall stage to be relevant, 
the evolution will proceed according to purely Ohmic field decay. 
The very existence of magnetars and of their presumable descendants, thermally emitting NSs with
$B\approx 10^{13}$ G, suggests that the fraction of NSs born with large magnetic
fields may be larger than expected.

\begin{acknowledgements} 
This work has been supported by the Spanish Ministerio 
de Ciencia y Tecnolog\'{\i}a grant AYA 2004-08067-C03-02. 
JAP is supported by a {\it Ram\'on y Cajal} contract from the Spanish MEC. 
\end{acknowledgements} 

\newpage 
\begin{appendix} 
\section{Induction equation in cylindrical coordinates.} 
In this appendix we use a different set of variables to derive 
the induction equation in cylindrical coordinates 
($R\equiv r\sin\theta, \varphi, z\equiv r\cos\theta$) and use the notation of Hollerbach \& R\"udiger (2002). 
Alternatively to our decomposition in Eq. (\ref{bfield}), one can in axial symmetry
simply work with the $\varphi$-components of the vector potential and of 
the magnetic field, 
\beq 
\vec{B}=\vec{B}_{\rm pol} + \vec{B}_{\rm tor} = 
\nabla \times (A_\varphi \vec{e}_\varphi) + B_\varphi \vec{e}_\varphi\;\;. 
\eeq 
In cylindrical coordinates, we can write explicitly the different 
components of the magnetic field and the current density: 
\beq 
\vec{B} &=& - \frac{\partial A_\varphi}{\partial z} \vec{e}_R 
+ \frac{1}{R}\frac{\partial (RA_\varphi)}{\partial R} \vec{e}_z 
+B_\varphi \vec{e}_\varphi \;\;,
\\ 
\frac{4 \pi}{c} \vec{J} = \nabla \times \vec{B} &=& - \frac{\partial B_\varphi}{\partial z} \vec{e}_R 
+ \frac{1}{R}\frac{\partial (RB_\varphi)}{\partial R} \vec{e}_z 
+ \hat{J}_\varphi \vec{e}_\varphi  \;\;,
\eeq 
where we have introduced the notation  
\beq 
\hat{J}_\varphi \equiv -\nabla^2 A_\varphi + \frac{A_\varphi}{R^2} ~.  
\eeq 
The induction equation in terms of this variables reads: 
\beq 
\frac{\partial A_\varphi}{\partial t} &=& - 
\frac{\hat{\tau}}{\hat{\sigma}}
\vec{e}_\varphi \cdot [(\nabla \times \vec{B}_{\rm tor}) \times \vec{B}_{\rm pol}]  
- \frac{\hat{J}_\varphi}{\hat{\sigma}} 
\nonumber \\ 
\frac{\partial B_\varphi}{\partial t} &=& 
\vec{e}_\varphi \cdot \nabla \times \left(\frac{\hat{\tau}}{\hat{\sigma}}
[(\nabla \times \vec{B}_{\rm pol}) \times \vec{B}_{\rm pol}  
 + (\nabla \times \vec{B}_{\rm tor}) \times \vec{B}_{\rm tor}] \right)
\nonumber \\ 
&+& \nabla \times \left[ \frac{1}{\hat{\sigma}} \nabla \times B_\varphi \right]\;\;. 
\eeq 
The term appearing in the r.h.s. of the first equation 
can be written as follows: 
\beq 
\vec{e}_\varphi \cdot [(\nabla \times \vec{B}_{\rm tor}) \times \vec{B}_{\rm pol}] = 
\frac{1}{R^2} \left[ \nabla (RA_\varphi) \times \nabla(RB_\varphi) \right]_\varphi~. 
\eeq 
The two terms inside the curl of the second equation are 
\beq 
(\nabla \times \vec{B}_{\rm tor}) \times \vec{B}_{\rm tor} &=& 
- \frac{B_\varphi}{R} \nabla (RB_\varphi)  
\nonumber \\ 
(\nabla \times \vec{B}_{\rm tor}) \times \vec{B}_{\rm pol} &=& 
\frac{J_\varphi}{R} \nabla (RA_\varphi)  
\eeq 
Then, taking the curl and after some algebra, 
the two terms appearing in the equation for $B_\varphi$ can be written as follows: 
\beq 
\vec{e}_\varphi \cdot \nabla \times [(\nabla \times \vec{B}_{\rm tor}) \times \vec{B}_{\rm tor}] &=& 
\left[ \nabla (RB_\varphi) \times \nabla(B_\varphi/R) \right]_\phi 
\nonumber \\ 
&=& - \frac{2 B_\varphi}{R} \frac{\partial B_\varphi}{\partial z} 
\nonumber \\ 
\vec{e}_\varphi \cdot \nabla \times [(\nabla \times \vec{B}_{\rm pol}) \times \vec{B}_{\rm pol}] &=& 
- \left[ \nabla (RA_\varphi) \times \nabla(\hat{J}_\varphi/R) \right]_\phi\;\;, 
\eeq 
where the first one is used in the derivation of Eq. (\ref{Burgers}). 
\end{appendix} 

\bibliographystyle{aa}

\end{document}